\documentclass{desyproc}

\newcommand{\pt}{p_T}
\newcommand{\ptjet}{p_{ T}^{jet}}
\newcommand{\ptjetem}{p_{ T}^{jet,em}}
\newcommand{\etajet}{\eta^{jet}}

\newcommand{\rapjet}{y^{jet}}

\newcommand{\akt}{\rm{anti}-k_T}
\newcommand{\njet}{N^{jet}}

\newcommand{\pttrk}{\pt^{track}}
\newcommand{\etatrk}{\eta^{track}}
\newcommand{\mjj}{{m}^{jj}}
\newcommand{\phijj}{\Delta \phi^{jj}}

\usepackage{subfigure}
\usepackage{verbatim}

\begin{document}
\title{Observation of Energetic Jet Production in \\$pp$ Collisions at $\sqrt{s}=7~$TeV using the ATLAS \\Experiment at the LHC}
\author{{\slshape Eric Feng}  for the ATLAS Collaboration\\[1ex]
Enrico Fermi Institute, University of Chicago, 5640 S. Ellis Ave, Chicago, IL 60637, USA }

\contribID{xy}  %
\confID{1964}
\desyproc{DESY-PROC-2010-01}
\acronym{PLHC2010}
\doi            %

\maketitle

\begin{abstract}
We report the observation of energetic jet production in proton-proton collisions at $\sqrt{s}=7$~TeV, based on about 1~nb${}^{-1}$ of integrated luminosity collected by the ATLAS detector.  The $\akt$ algorithm is used to reconstruct jets with $\ptjet >30$~GeV 
and $|\rapjet|<2.8$.  Jets with $\ptjet$ up to $\sim 500$~GeV and events with dijet mass up to $\mjj \sim 1$~TeV are observed.  The jet shapes and charged particle flow confirm that the observed jet signal corresponds to collimated flows of particles in the final state.
\end{abstract}

\section{Introduction}
The observation of energetic jets produced in $pp$ collisions at $\sqrt{s}=7$~TeV is reported, using about $1$~nb${}^{-1}$ of data collected by the ATLAS experiment.  Kinematic distributions in inclusive jet and dijet production are presented, together with internal jet structure and charged particle flow in the event.  No attempt is made to correct the measurements for detector effects or account for systematic uncertainties.

These measurements were performed using the ATLAS detector, which is a general purpose, hermetic detector described in detail elsewhere~\cite{atlas}.  The ATLAS tracking system covers the pseudorapidity range $|\eta|<2.5$, while the electromagnetic and hadronic calorimeters cover $|\eta|<4.9$.  The data are compared to PYTHIA 6.4.21~\cite{pythia}, which is based on leading order $2 \rightarrow 2$ perturbative QCD matrix elements plus parton shower.  The Monte Carlo (MC) uses a set of tuned parameters denoted as ATLAS MC09~\cite{mc09} along with MRST LO${}^{*}$ parton density functions~\cite{mrst}, and the full ATLAS detector response is modeled in GEANT4~\cite{geant}.

\section{Event selection}
The data were collected during the first LHC runs at $\sqrt{s} = 7$~TeV 
in March and April 2010.
Events were triggered by requiring at least one hit from minimum bias trigger scintillators (MBTS)~\cite{mbts} that cover 2.09~$<|\eta|<$~3.84.
The events are required to  
have a reconstructed primary vertex with 
a $z$-position within 10~cm
of the detector center in order to suppress beam-related backgrounds and cosmic rays.
Additional quality criteria are also applied to ensure that jets are not produced by 
single noisy calorimeter cells or problematic detector regions~\cite{DQconf}.

\section{Jet reconstruction}
Jets are identified using the $\akt$ jet algorithm~\cite{akt} with distance parameter $R=0.6$ by performing four-momentum recombination on topological clusters~\cite{cluster}.  These clusters are seeded by calorimeter cells with $|E_{\rm cell}|>4\sigma$ above the cell energy noise.  All directly neighbouring cells are added, then neighbors of neighbours are iteratively added for all cells with signals above a secondary threshold $|E_{\rm cell}|>2\sigma$.
The measured jet transverse momentum $\ptjetem$, as determined at the electromagnetic scale~\footnote{The electromagnetic scale is the appropriate scale for the reconstruction of the energy deposited by electrons or photons in the calorimeter.}, systematically underestimates that of the hadron-level jet due to calorimeter non-compensation and dead material.  Consequently an average correction $C(\ptjetem,\etajet)$, determined as a function of $\ptjetem$ and $|\etajet|$ from MC simulation, is applied to obtain the corrected $\ptjet$.
No attempt is made to unfold the effects of the finite detector resolution.

Events are required to have at least one jet with (corrected) $\ptjet > 30$~GeV and $|\rapjet| < 2.8$.  
Preliminary studies indicate that for jets with $|\rapjet | < 2.8$, the relative response of the 
calorimeter to jets in different rapidity regions is correctly modeled by MC to within $\pm 5 \%$.  A first determination of the energy scale for jets, using in-situ isolated tracks and
calorimeter $(E/p)$ measurements and test beam results,  
establishes an absolute jet energy scale uncertainty of about $\pm 7 \%$.

\section{Results}
\subsection{Inclusive Jet Production}
\begin{figure}[tbh]
\begin{center}
\subfigure[
\label{fig:mult}]{
\includegraphics[width=0.32\textwidth]{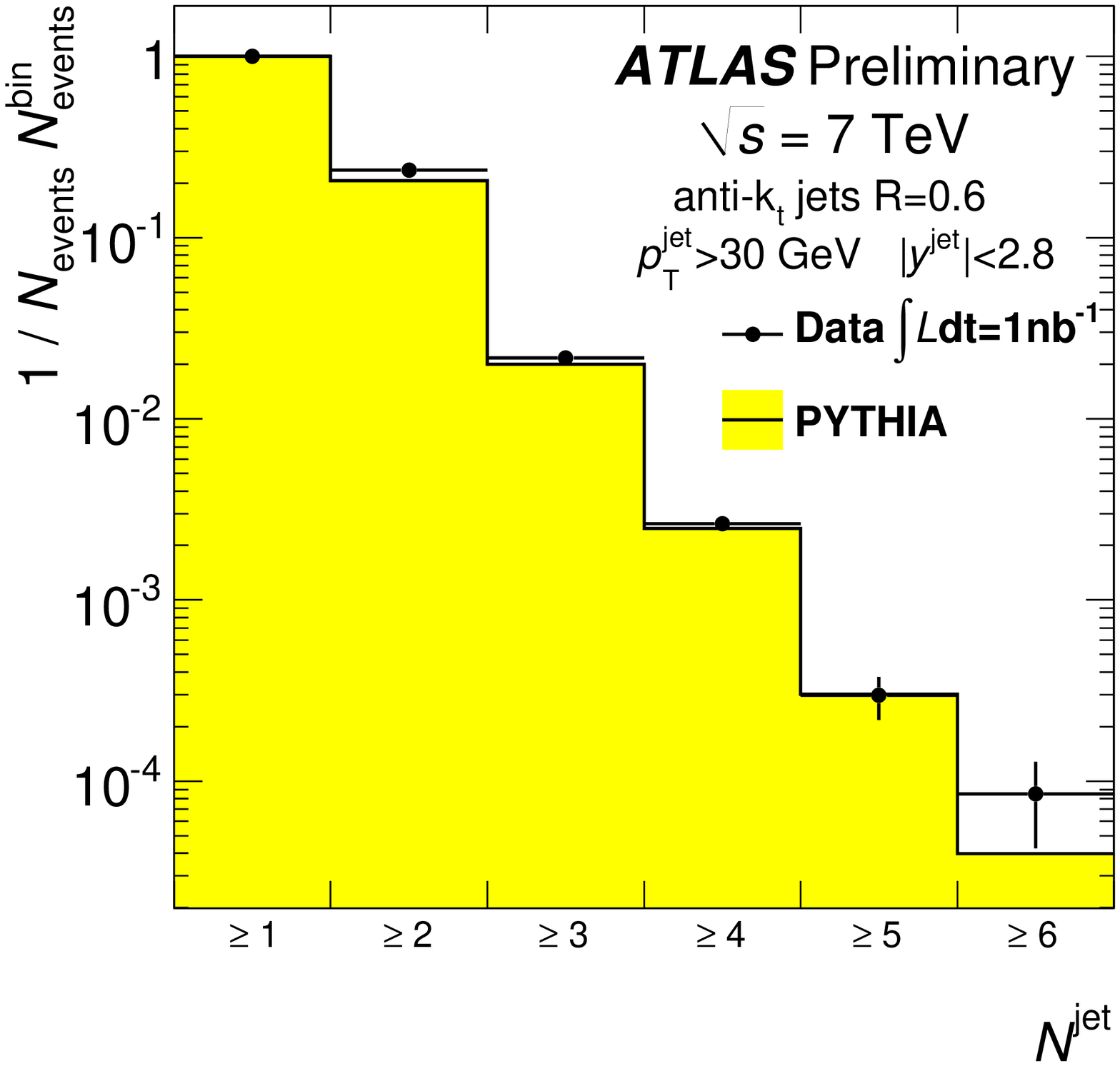}}
\subfigure[
\label{fig:pt}]{
\includegraphics[width=0.32\textwidth]{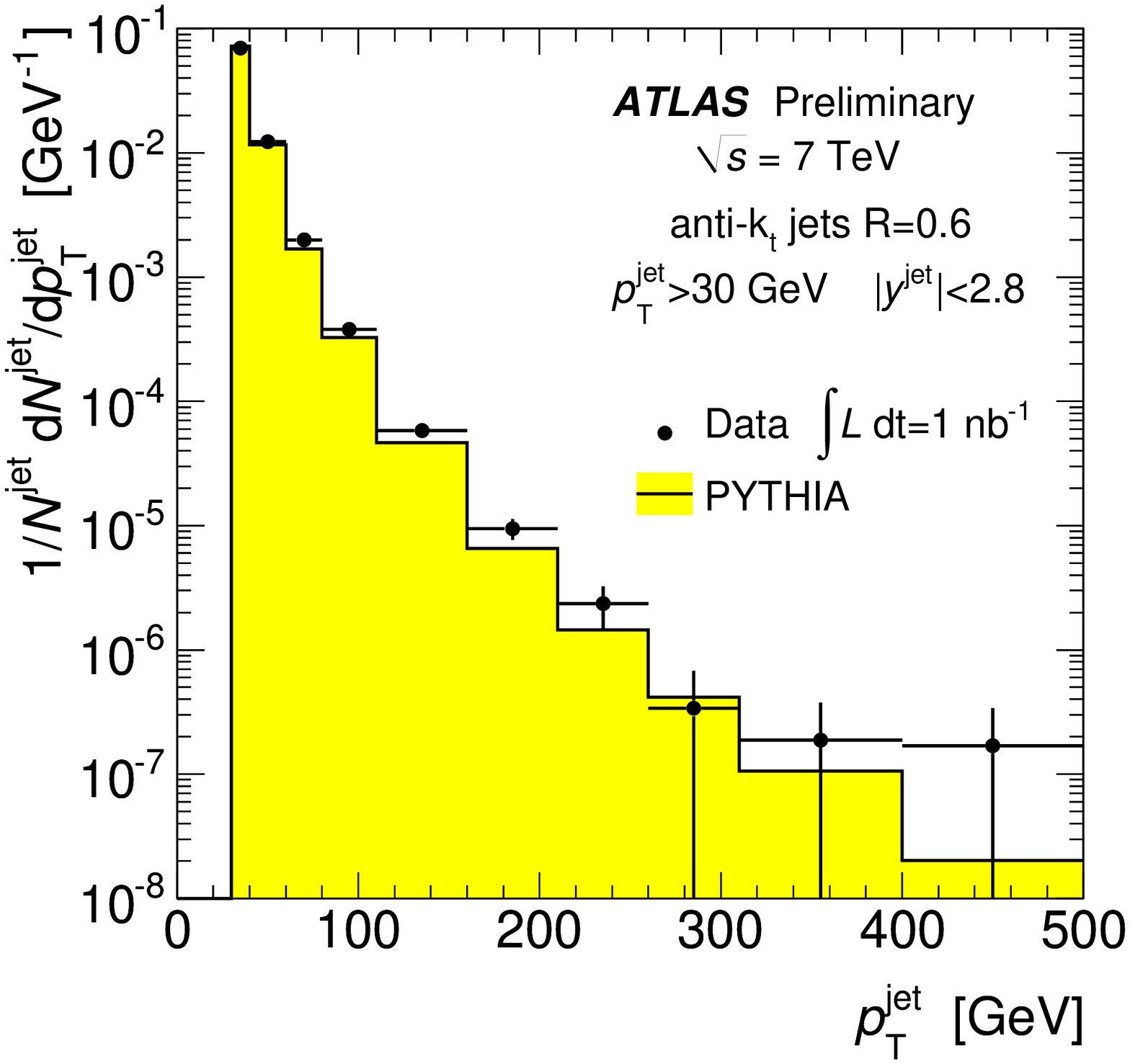}}
\subfigure[
\label{fig:rapidity}]{
\includegraphics[width=0.32\textwidth]{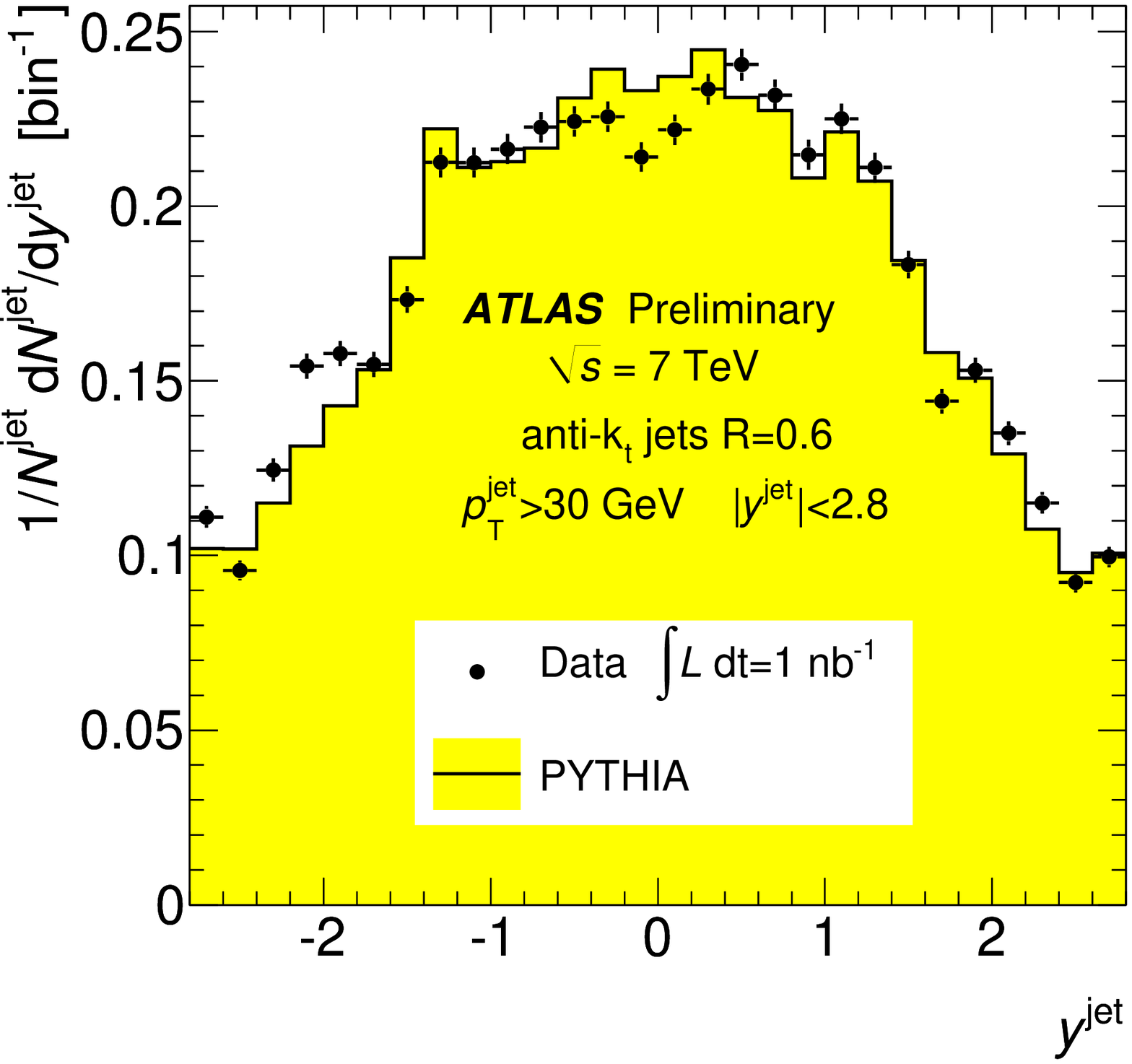}}
\end{center}
\vspace{-0.2 cm}
\caption{\small
Observed inclusive jet multiplicity $(\geq \njet)$ distribution (left), $\ptjet$ (center), and $\rapjet$ (right).  The distributions are normalized to unity and only statistical uncertainties are included.
}
\label{fig:mult_pt_rapidity}
\end{figure}

\noindent
Figure~\ref{fig:mult_pt_rapidity} presents the multiplicity, transverse momentum, and rapidity distributions for all jets with $\ptjet > 30$~GeV and $|\rapjet| < 2.8$.
Events with six jets in the final state are observed, and
jets are observed with $\ptjet$ up to $\sim$500~GeV.  The MC provides a reasonable description of the distributions, but still shows some deficiencies in the observed jet rapidity distribution.
\begin{figure}[tbh]
\begin{center}
\subfigure[
\label{fig:mjj}]{
\includegraphics[width=0.32\textwidth]{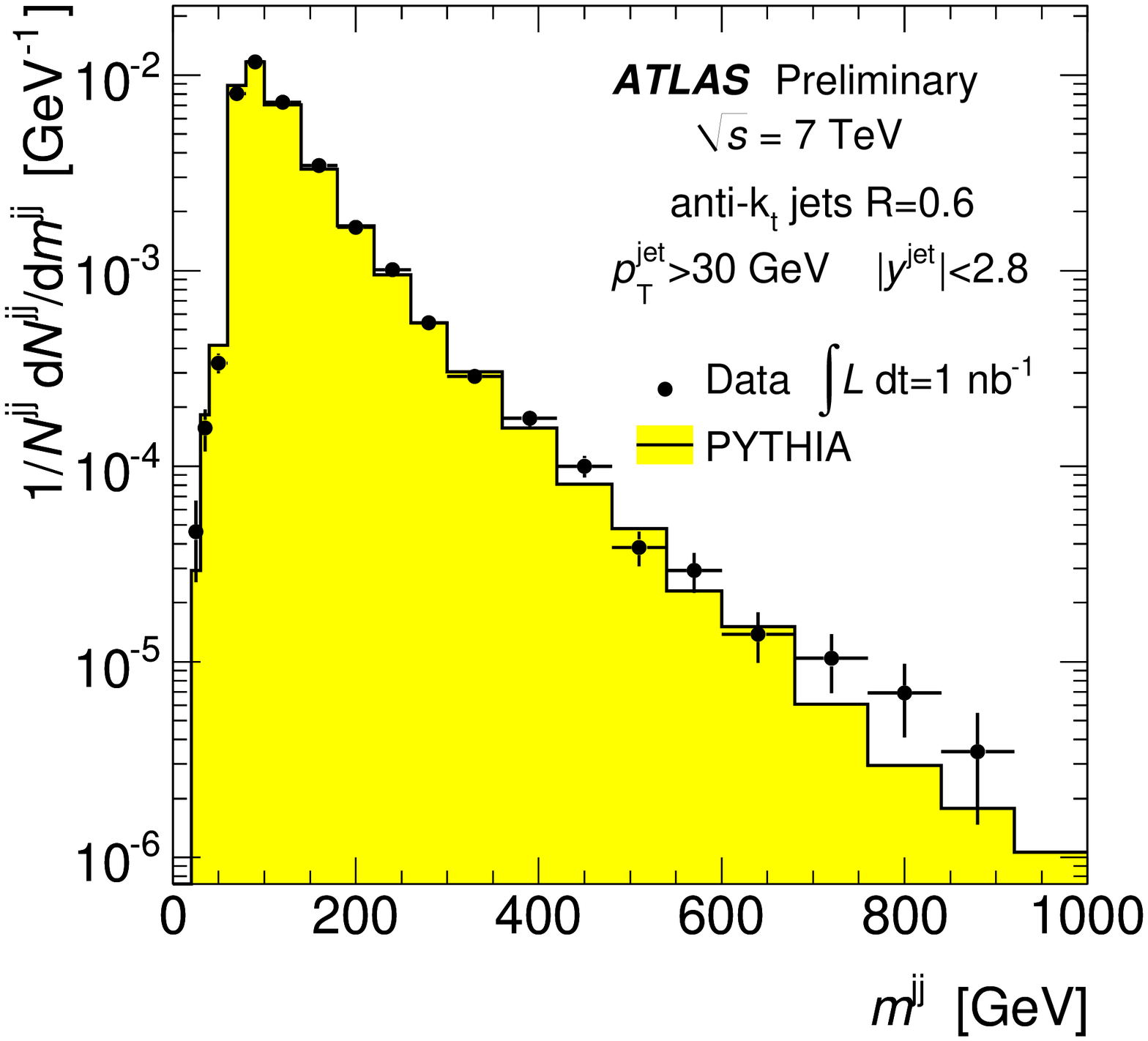}}
\subfigure[
\label{fig:dphi}]{
\includegraphics[width=0.32\textwidth]{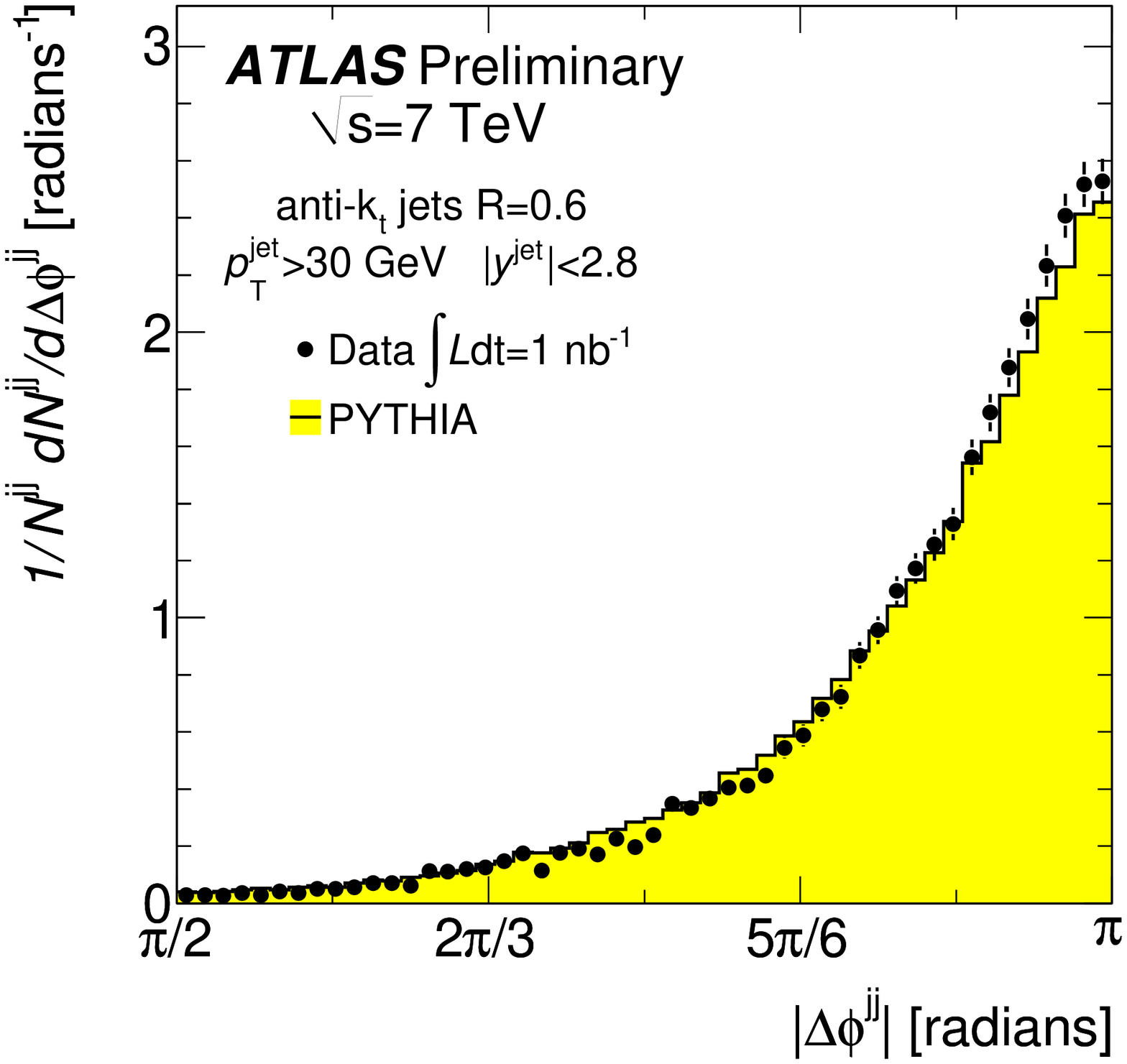}}
\end{center}
\vspace{-0.2 cm}
\caption{\small
Observed $\mjj$ (left) and $|\phijj|$ (right) distributions in inclusive dijet events. Only statistical uncertainties are included and the distributions are normalized to unity.
}
\label{fig:dphi_mjj}
\end{figure}
The invariant mass $\mjj$ of the two leading jets and their azimuthal angular separation $|\phijj|$ are presented in 
Fig.~\ref{fig:dphi_mjj}.
The shape of the $\mjj$ distribution at low mass reflects the limited phase space due to the thresholds applied on $\ptjet$ and $\rapjet$.  Above that, the observed spectrum decreases with increasing $\mjj$ up to a dijet mass around 1~TeV.
The observed $|\phijj|$ distribution strongly peaks at $|\phijj| \sim \pi$,  
indicating a dominant back-to-back dijet configuration in the final state.  The shapes of the dijet mass spectrum and $|\phijj|$ distribution are described by MC simulation, though the MC underestimates the data at large $|\phijj|$. 

\subsection{Jet Shapes and Charged Particle Flow}

\begin{figure}[tbh]
\begin{center}
\subfigure[
\label{fig:shapes_sketch}]{
\includegraphics[width=0.12\linewidth]{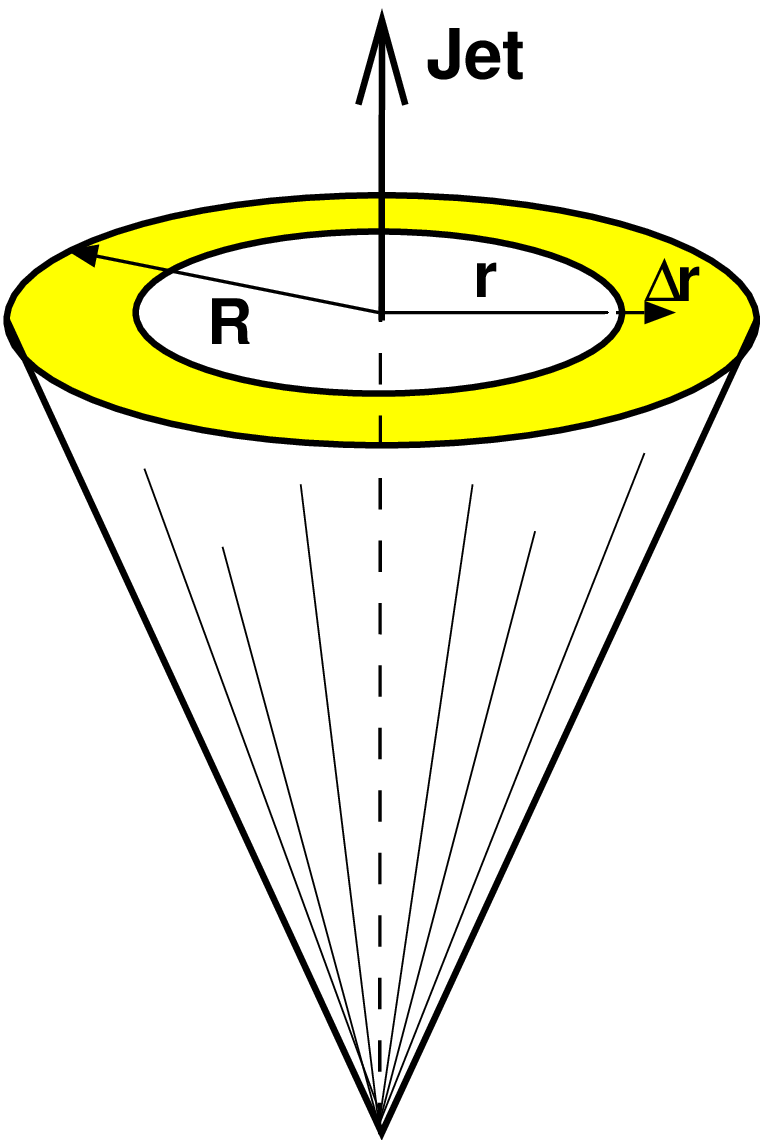}}
\hspace*{0.10\textwidth}
\subfigure[
\label{fig:flows_sketch}]{
\includegraphics[width=0.35\linewidth]{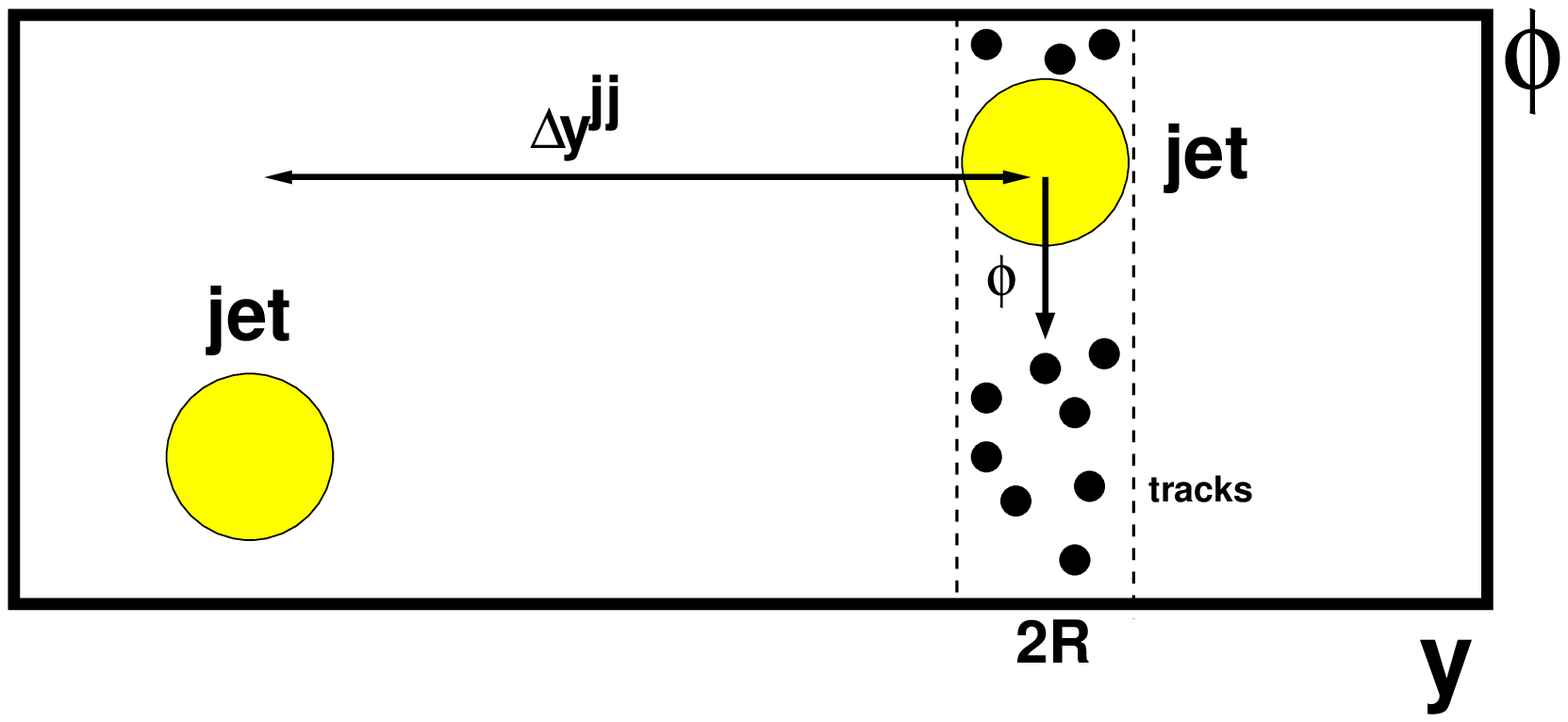}}
\caption{
\small
Sketch of the jet shape as a function of the distance to the jet axis (left), and sketch of the particle flow as a function of the distance to the jet axis in the azimuthal direction (right).}
\vspace{-0.2 cm}
\label{fig:sketches}
\end{center}
\end{figure}

\begin{figure}[tbh]
\begin{center}
\includegraphics[width=0.50\linewidth]{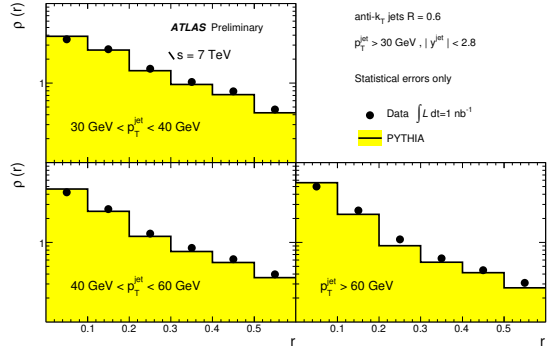}
\vspace{-0.2 cm}
\caption{
\small
Observed differential jet shapes $\rho(r)$ in different regions of jet transverse momentum.
}
\label{fig:shapes}
\end{center}
\end{figure}

The transverse momentum distribution inside the jet and the charged particle flow around the jet, which are illustrated schematically in Fig.~\ref{fig:sketches}, are studied in order to 
test our quantitative understanding of the jet properties.
The differential jet shape illustrated in Fig.~\ref{fig:shapes_sketch} is
defined as the average fraction of jet transverse momentum density within an annulus spanning ${r \pm \Delta r/2}$ around the jet axis:

\begin{equation}
\rho (r) = \frac{1}{\Delta \ r}\frac{1}{\njet}\sum_{\rm jets}\frac{p_{T}(r-\Delta \ r/2,r+\Delta \ r/2)}{p_{T}(0,R)},~0 \le r \le R
\end{equation}

\noindent
where $ p_{T}$ denotes the scalar sum of the transverse momentum of the 
calorimeter clusters in a given 
annulus, $\njet$ is the number of jets, $R=0.6$, and $\Delta r = 0.1$ are used. The observed jet shapes are presented in Fig.~\ref{fig:shapes} for jets with $\ptjet >30$~GeV and $|\rapjet|<2.8$ in different regions of jet transverse momentum. The distributions
peak at low $r$, indicating the presence of a collimated flow of particles around the jet axis. 
The measurements are reasonably well described by the PYTHIA MC, which tends to produce slightly narrower jets than the data.
\begin{figure}[tbh]
\begin{center}
\includegraphics[width=0.55\linewidth]{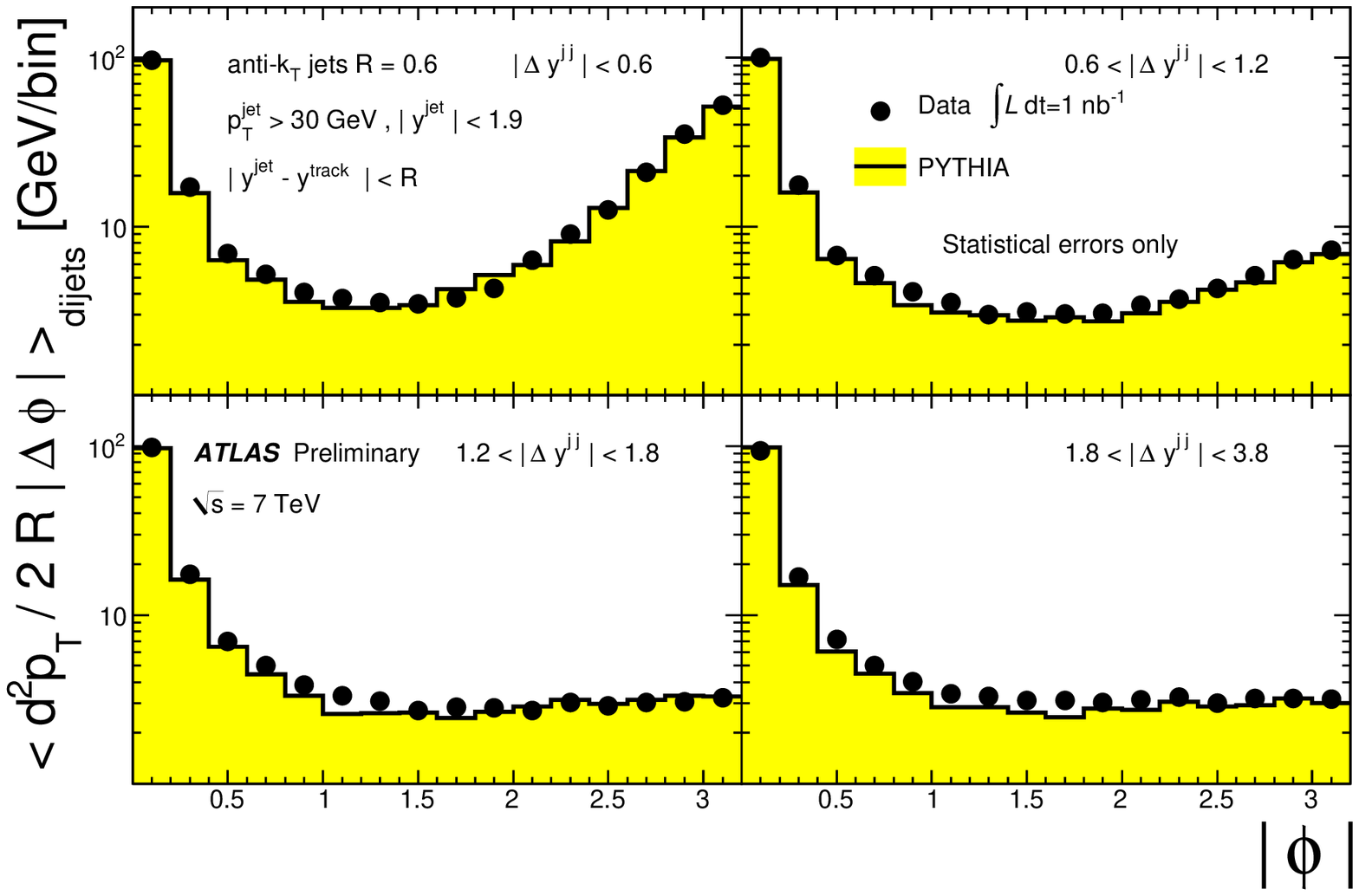}
\vspace{-0.2 cm}
\caption{
\small
Observed charged particle flow in inclusive dijet events as a function of $|\phi|$ with respect to the jet direction and the rapidity separation between the two leading jets.
}
\label{fig:flows}
\end{center}
\end{figure}

The charged particle flow around the jet provides an independent track-based technique to confirm the calorimeter-based jet shapes and to understand the final state topology.  The hadronic activity out of the jet cone is studied in inclusive dijet events using tracks, which are selected as in~\cite{MBpaper} using $\pttrk > 500$~MeV and 
$|\etatrk|<2.5$. The average transverse momentum is measured as a function of the azimuthal distance to the jet axis as illustrated in Fig.~\ref{fig:flows_sketch} and is defined as:

\begin{equation}
<\frac{  d^2 p_{T}}{|d \phi| d y }>_{jets} = 
\frac{1}{ 2 R |\Delta \phi| }\frac{1}{\njet}\sum_{jets} {  p_{T}(|\phi-\Delta \phi/2|,|\phi+\Delta \phi/2|)}, {\rm with}~0 \le |\phi| \le \pi,
\label{eq:eflows}
\end{equation}

\noindent
where  $p_{T}(|\phi-\Delta \phi/2|,|\phi+\Delta \phi/2|)$ is the scalar sum of the transverse momentum of the 
tracks at a given distance $\phi$ to the jet, and bins of $\Delta \phi = 0.2$ are used.  Only tracks within the rapidity range spanned by the jet cone are included.  The jet is required to have $\ptjet > 30$~GeV and $|\rapjet|<1.9$ as determined by the tracking coverage of $|\etatrk| < 2.5$.  This is performed as a function of the rapidity separation between the two leading jets $|\Delta y^{jj}|$.
In Fig.~\ref{fig:flows},
for $|\Delta y^{jj}| < 0.6$ the presence of two collimated jets of tracks at $|\phi| \sim 0$ and $|\phi| \sim \pi $ is observed as expected.  For $|\Delta y^{jj}| > 1.2$, the jet structure for $|\phi| < 0.6$ is followed by a plateau of remaining hadronic activity as $|\phi|$ increases.  The PYTHIA MC provides a reasonable description of the data, but slightly underestimates the hadronic activity away from the jet direction.

\section{Summary}
We have reported the observation of energetic jet production in $pp$ collisions at $\sqrt{s}=7$~TeV, based on about 1~nb${}^{-1}$ of data collected by the ATLAS detector.  The $\akt$ algorithm is used to reconstruct jets with $\ptjet >30$~GeV and $|\rapjet|<2.8$ from calorimeter energy clusters.  Jets with $\ptjet$ up to $\sim 500$~GeV and events with dijet mass up to $\mjj \sim 1$~TeV are observed.  The jet shapes and charged particle flow confirm that the observed jet signal corresponds to collimated flows of particles in the final state.

\section{Acknowledgments}

The author is grateful for support from the NSF US LHC Graduate Student Support Award.

\section{Bibliography}

\begin{footnotesize}

\end{footnotesize}

\end{document}